# A neuroeconomic theory of rational addiction and nonlinear time-perception.

Taiki Takahashi[1]

[1]Direct all correspondence to Taiki Takahashi (Email taikitakahashi@gmail.com)
Department of Behavioral Science, Faculty of Letters, Hokkaido University
N.10, W.7, Kita-ku, Sapporo, 060-0810, Japan
TEL +81-11-706-3057, FAX +81-11-706-3066.

**Acknowledgements:** The research reported in this paper was supported by a grant from the Grant- in-Aid for Scientific Research (Global COE) from the Ministry of Education, Culture, Sports, Science and Technology of Japan.

Abstract

Neuroeconomic conditions for "rational addiction" (Becker and Murphy, 1988) have been unknown. This paper derived the conditions for "rational addiction" by utilizing a nonlinear time-perception theory of "hyperbolic" discounting, which is mathematically equivalent to the q-exponential intertemporal choice model based on Tsallis' statistics. It is shown that (i) Arrow-Pratt measure for temporal cognition corresponds to the degree of irrationality (i.e., Prelec's "decreasing impatience" parameter of temporal discounting) and (ii) rationality in addicts is controlled by a nondimensionalization parameter of the logarithmic time-perception function. Furthermore, the present theory illustrates the possibility that addictive drugs increase impulsivity via dopaminergic neuroadaptation without increasing irrationality. Future directions in the application of the model to studies in neuroeconomics are discussed.

1. **Introduction:**
Delay discounting in intertemporal choice refers to the devaluation of a delayed reward compared to the value of an immediate reward (Bickel & Marsch, 2001; Takahashi, 2009). Impulsivity in intertemporal choice (referred to as "impatience" in behavioral economics) is defined as strong preference of smaller but more immediate rewards over larger but more delayed ones. Economists Becker and Murphy (1988) proposed a theory of rational addiction which associates impulsivity in intertemporal choice and addictive behavior. For instance, heroin addicts prefer "sooner but (objectively) smaller" rewards (i.e., pleasures from drug intake) over "later but (objectively) larger" rewards (e.g., a long lifespan, and a healthy and rich elderly life). Consistent with the proposed association between impatience and addiction in the economic theory of rational addiction (Becker and Murphy, 1988), our neuroeconomic study have demonstrated that daily nicotine intake by smokers are associated with strong preference for smaller sooner rewards over larger later ones (Ohmura et al., 2005). Agent A who prefers "one glass of caipirinha available one year later" over "two glasses of caipirinha available [one year plus one week] later" is more impulsive (impatient) than agent B who prefers "two glasses of caipirinha available [one year plus one week] later" over "one glass of caipirinha available one year later". In this example 1, most people may behave like the patient agent B. It is to be noted that both impatient agent A and patient agent B may be rational, because, in this example 1 alone, there is no inconsistency even in impatient agent A's behavior. Suppose the next intertemporal choice example 2. There are two options: "one glass of caipirinha available today" and "two glasses of caipirinha available one week later". In example 2, most people who planned to choose the larger later option in example 1 simultaneously tend to prefer "one glass of caipirinha available today" over "two glasses of caipirinha available one week later". The combination of these two intertemproal choices in example 1 (choosing the larger later) and example 2 (choosing the smaller sooner) is time-inconsistent (irrational). Namely, agents who prefer larger later rewards in the distant future, but prefer smaller sooner rewards in the near future are dynamically inconsistent (irrational), because their preferences reverse as time passes (Laibson, 1997). Economists Becker and Murphy (1988)'s theory of rational addiction assumes addicts are "rational" (in time-consistency and maximization of the summed temporally-discounted utility over her lifespan) and impatient, resulting in the consumption of harmful drugs at the cost of healthy life in the old age. However, it is still possible that addicts are not only more impatient but also more irrational (i.e., more time-inconsistent) than non-drug-dependent subjects, as economists Gruber and Koszegi (2001) questioned. Therefore, it is important to examine neuroeconomic psychophysical, and biophysical conditions under which addicts are rational (even though they are impatient), in order to develop neuroeconomic theory of addiction and extend the frameworks of econophysics into time-inconsistency related to psychophysics of temporal cognition.

I derive, in this paper, conditions on parameters in neuroeconomic theory of intertemporal choice which incorporates psychophysics of time-perception (Kim and Zauberman, 2009; Takahashi, 2005; Takahashi, 2006; Takahashi et al., 2008; Zauberman et al., 2009), and is mathematically equivalent to the q-exponential discount model based on Tsallis' statistics (Cajueiro, 2005; Takahashi et al., 2007). Notably, the q-exponential function is a well-studied function in a deformed algebra developed in Tsallis' non-extensive thermodynamics (Tsallis, 1994).

This paper is organized in the following manner. In Section 2, I briefly introduce the discount models in relation to time-inconsistency (irrationality) and time-perception, and their relations to the q-exponential discounting. In Section 3, I derive the conditions on the parameters in the introduced discount models and give neuroeconomic and biophysical and psychophysical interpretations to the conditions on the parameters. In Section 4, some conclusions from this study and future study directions by utilizing the present theoretical framework in neuroeconomics of addiction are discussed.

2. **Rationality in intertemporal choice and psychophysics of time-perception**
2.1 **Discount function based on Tsallis' statistics**

Rational (i.e., time-consistent) temporal discounting which has been assumed in neoclassical economic theory including Becker and Murphy's theory of rational addiction (1988), follows the exponential discount function:

$$V(D) = V(0)exp(-k_e D) \qquad (1)$$

where $V(D)$ is the subjective value of a reward received at delay D, $V(0)$ is the value of an immediate reward, and D is the length of delay until the delivery of reward. The free parameter $k_e$ is an index of the degree to which the delayed reward is discounted, i.e., larger $k_e$ values correspond to steeper delay discounting. In exponential discounting, there is no time-inconsistency because a time discount rate defined as $-V(D)'/V(D)$ is constant ($=k_e$) over time. However, as introduced above, human (and animal) subjects mostly display preference reversals over time, because the following hyperbolic discount equation better fits behavioral data compared to the exponential discount

function:

$$V(D) = V(0)/(1+k_h D). \quad (2)$$

where large $k_h$ values again correspond to steep discounting. It is important to note that, in hyperbolic discounting, the discounting rate $= -V'/V = k_h/(1+k_h D)$ (a hyperbolic discount rate) is a decreasing function of delay, resulting in preference reversal as time passes. In order to quantify the two distinct type of behavioral tendencies in intertemporal choice (i.e., irrationality and impatience), recent studies in econophysics and neuroeconomics have introduced (Cajueiro, 2005) and examined (Takahashi et al., 2007) the following q-exponential discount function based on Tsallis' statistics.:

$$V_q(D) = V_q(0)/\exp_q(k_q D) = V_q(0)/[1+(1-q)k_q D]^{1/(1-q)} \quad (3)$$

where $V_q(D)$ is the subjective value of a reward obtained at delay D and $k_q$ is a parameter of impulsivity at delay $D=0$ (q-exponential discount rate at delay D=0). When $q=0$, equation 3 is the same as a hyperbolic discount function (equation 2), while $q \to 1$, is the same as an exponential discount function (equation 1). We have previously demonstrated that this q-parameter can be used to assess the deviation of human intertemporal choice from exponential discounting in behavioral studies (Takahashi et al., 2007).

**2.2 Role of psychophysics of time-perception in intertemporal choice**

It has been proposed, based on neurochemical and psychophysical findings, that incorporating psychophysical effects on time-perception such as Weber-Fechner law (i.e. logarithmic time-perception) into exponential discounting may be capable of describing empirically observed irrationality in intertemporal choice (Takahashi, 2005). Because in logarithmic time-perception, subjective time-duration $\tau$ (psychological time) is expressed as:

$$\tau(D) = a \ln(1+bD), \quad (4)$$

exponential discounting with the subjective delay $\tau$ (exponential discounting with logarithmic time-perception) is expressed as:

$$V(D) = V(0)\exp(-k\tau)$$
$$= V(0)\exp(-ka \ln(1+bD))$$

$$= V(0)/(1+bD)^g \qquad (5)$$

where *V(0)* is the value of the immediate reward, D is an objective/physical delay length and b and *g=ka* are free parameters. In equation 4, parameter b is a nondimensionalization coefficient of physical time with a dimension of $[T^{-1}]$, while parameter a is physically dimensionless (note that psychological time is dimensionless in the physical system). Also, parameter k in equation 5 is a physically-dimensionless "intrinsic impulsivity (impatience)" parameter in that agents with large k is impulsive (impatient) in intertemporal choice with respect to her psychological time τ. Note also that, as one can see from equation 3 and 5, by utilizing relationships *q=(g-1)/g* and $k_q=bg$, we obtain parameters in the q-exponential discount model from those in the logarithmic-time exponential discount function (equation 5). Furthermore, recent studies in behavioral economics confirmed the logarithmic time-perception theory of irrational discounting (Takahashi et al., 2008; Kim and Zauberman, 2009; Zauberman et al., 2009).

### 2.3 Parameters of impatience and irrationality in intertemporal choice

Impulsivity in intertemporal choice (referred to as "impatience" in behavioral economics) at delay D is parameterized as the discount rate: DR(D)= - V'(D)/V(D). The time-dependency of the discount rate is DR'(D). In order to define the degree of "preference reversal" over time in intertemporal choice, behavioral economist Prelec (2004) axiomatically analyzed intertemporal choice and proposed the following "decreasing impatience" parameter:

$$DI(D) = -\frac{[\ln \varphi(D)]''}{[\ln \varphi(D)]'}, \qquad (6)$$

where $\varphi(D)$ is a time-dependent part of discount function: V(D)/V(0). By utilizing the definition of the discount rate DR(D)= - V'(D)/V(D)= - [$\varphi(D)$]'/[$\varphi(D)$], we can see that

$$DI(D) = -\frac{DR'(D)}{DR(D)}. \qquad (7)$$

Therefore, Prelec's DI can be interpreted as a time-decay rate of the discount rate, which is zero for exponential discounting and positive for irrational intertemporal choice associated with preference reversal over time. This study is the first to give this simple interpretation for Prelec's DI (i.e., a time-decay rate of the discount rate). Because the degree of irrationality in intertemporal choice (the degree of "preference reversal" over time) can be parametrized with this DI parameter, we can utilize the DI parameter to derive conditions of rational addiction on parameters in the exponential discounting model with logarithmic time-perception (equation 5). It is to be noted that economists Arrow (1965) and Pratt (1964) invented the nonlinearity (concavity) parameter of the utility function, in order to parametrize the degree of risk-aversion based on von Neumann-Morgenstern's expected utility theory (von Neumann and Morgenstern, 1947). Prelec's DI (2004) corresponds to Arrow-Pratt measure of concavity of the logged discount functions.

3. **Conditions for rational addiction**
**3.1 Rationality in intertemporal choice and nonlinear time-perception**
Let us now return to the theory of rational addiction (Becker and Murphy, 1988). The rational addiction theory states that addicts (e.g. habitual smokers and alcoholics) are more impatient than non-drug-dependent subjects, but not more irrational than non-drug-dependent subjects in utility maximization over time. To examine how this is possible, by adopting the experimentally-confirmed theory of irrational time-discounting (i.e., nonlinear time-perception theory of temporal discounting, equation 5) is the objective of this section. By applying the definition of Prelec's parameter of irrationality in intertemporal choice (equation 6) to the following exponential discount model with general nonlinear time-perception:

$$V(D) = V(0)\exp(-k\tau(D)),  \qquad (8)$$

we obtain the following expression of Prelec's DI in terms of $\tau(D)$:

$$\mathrm{DI}(D) = -\frac{\tau''(D)}{\tau'(D)},  \qquad (9)$$

which indicates that irrationality in intertemporal choice is equal to Arrow-Pratt measure of nonlinearity (concavity) of the general psychophysical function of time-perception. This is the first study to demonstrate that nonlinearity in time-perception, which is parametrized with Arrow-Pratt measure of psychological time, equals Prelec's DI.

Next, we insert the logarithmic time-perception (equation 4) to equation 9 to

obtain

$$\text{DI(D) (for log-time theory of discounting)} = \frac{b}{1+bD}, \tag{10}$$

which demonstrates that irrationality in intertemporal choice is only related to the nondimensionalization parameter b, but unrelated to intrinsic impatience $k$ and the dimensionless coefficient of psychological time (parameter *a*) in equation 4. Furthermore, a partial derivative of DI in terms of b is

$$\frac{\partial}{\partial b} DI = \frac{1}{(1+bD)^2} > 0, \tag{11}$$

indicating that irrationality is enhanced as the nondimensionalization coefficient of psychological time in equation 4 increases. Let us then examine how the discount rate depends on parameters in exponential discounting with psychological time.

$$\begin{aligned} DR(D) &= -V'/V \\ &= k\,\tau'(D) \\ &= \frac{kab}{1+bD} \\ &= ka[DI(D)], \end{aligned} \tag{12}$$

indicating that the discount rate of equation 5 increase as any of intrinsic impatience k, psychophysical time-perception parameters a and b increases (note that DI(D) is an increasing function of b, as indicated in equation 11). Therefore, it can be concluded that if addictive drugs do not increase the nonlinearity of psychophysical function of time-perception, but potentiate intrinsic impatience and prolong psychological time (possibly via dopaminergic neuroadaptations in the brain regions such as the striatum), drug addicts may still be as rational as non-drug-dependent subjects, even if they are impatient, which is consistent with Becker and Murphy's theory of rational addiction (Becker and Murphy, 1988).

### 3.2 Rationality in q-exponential discounting based on Tsallis' statistics

In order to utilize the present framework in Tsallis' statistics-based econophysics, I here denote the parameters of impatience and irrationality (Prelec's DI) in the q-exponential discount model based on Tsallis' statistics (equation 3):

$$DR_q(D) = \frac{k_q}{1 + k_q(1-q)D}. \tag{13}$$

$$DI_q(D) = \frac{k_q(1-q)}{1 + k_q(1-q)D}. \tag{14}$$

Here we can see that when $q \to 1$, $DR_1$ is independent of D and $DI_1=0$, corresponding to rational discounting (exponential discounting). Also, by utilizing relationships $q=(g-1)/g$, $k_q=bg$, and $g=ka$, we obtain equation 10 again. Because the q-exponential function has been developed in Tsallis' statistics-based econophysics (Anteneodo et al., 2002), future statistical physical and econophysical studies on temporal discounting can readily utilize these equations 13 and 14 and then translate obtained findings into the logarithmic time-perception theory of irrational time-discounting, which has both psychophysical interpretations and neuroeconomic importance.

4. **Conclusions and implications for neuroeconomics and econophysics**

Recent human neuropsychopharmacological studies revealed that neuropsychiatric patients such as ADHD (attention-deficit hyperactivity disorder)'s temporal discounting and time-perception may be modulated by dopamine systems in the brain (Rubia et al., 2009). Also, time-perception is related to dopaminergic drug-related habit formation (Williamson et al., 2008) and impaired time-perception is related to D2 (a subtype of dopamine receptors) functioning in the striatum (Ward et al., 2009). These findings indicate the important roles of dopamine in time-discounting and time-perception. Furthermore, we have previously shown that depressed patients (known to have impairment in serotonin activity in the brain) are more time-inconsistent than healthy subjects (Takahashi et al., 2008). Therefore, it can be expected that serotonin activity modulates the nondimensionalization parameter b of the psychophysical time-perception function (equation 4), resulting in exaggerated irrationality in temporal discounting. These points should be studied in future neuroeconomic studies.

Finally, it is known that magnitude and sign (i.e. gain/loss) of outcomes affects discount rates (Estle et al., 2006). It is possible these effects occur via the psychophysical effects on time-perception. For instance, while waiting for a larger reward, subjects do not strongly feel the passage of physical time, resulting in the "magnitude effect" on temporal discounting (i.e., larger rewards are less steeply time-discounted than smaller rewards). If this is the case, future studies should examine whether the effects of magnitude on temporal cognition affect nonlinearity of time-perception (indicated by parameter b in equation 4).